\documentclass[12pt,epsf]{article}

\usepackage{amsmath,amssymb,graphicx}

\newcommand{\beq}{\begin{eqnarray}}% can be used as {equation} or  {eqnarray}
\newcommand{\eeq}{\end{eqnarray}}

\title{
{\huge \bf Continuum Superpartners} \vspace*{2mm} {\huge \bf from
Supersymmetric Unparticles} \vspace*{0.5cm}
\author{\textbf{Haiying Cai, Hsin-Chia Cheng, Anibal D.~Medina}\\
 \textbf{and John Terning}\\
\\
\normalsize\emph{Department of Physics, University of California,
One Shields Ave. Davis, CA 95616, USA}\\}}

\date{\today}
\begin{document}
\setcounter{page}{0} \maketitle
\thispagestyle{empty}
\vspace*{0.5cm} \maketitle %\begin{center}
\begin{abstract}
We examine supersymmetric theories with approximately conformal
sectors.  Without an IR cutoff the theory has a continuum of
modes, which are often referred to as ``unparticles." Making use
of the AdS/CFT correspondence we find that in the presence of a
soft-wall, a gap in the spectrum can arise, separating the
zero-modes from the continuum modes. In some cases there are also
discrete massive levels in the gap. We also show  that when
supersymmetry is broken the superpartner of a quark or lepton may
simply be a bosonic continuum above a gap.  Such extensions of the
standard model have novel signatures at the LHC.
\end{abstract}

\thispagestyle{empty}

\newpage

\setcounter{page}{1}

\section{Introduction}

Supersymmetry (SUSY) is the most studied extension of the standard
model, but it seems that there are still unexplored possibilities
for observable signatures a the LHC. Here we will explore the case
where the SUSY theory has an approximate conformal symmetry.
Georgi has proposed \cite{Georgi}  using ``unparticles" as an
efficient means to calculate processes in such models.  To date
there have been many variations on the unparticle  idea (hidden
valleys \cite{Strassler:2006im}, quirks \cite{quirks}, massive
unparticles \cite{Fox:2007sy}, colored unparticles
\cite{coloredunparticles}, the Unhiggs \cite{unhiggs,Falkowski},
etc.) and a variety of possible collider signatures
\cite{Cheung:2007ap} have been studied.  Here we will explore
unparticles in the context  of SUSY. We would like to see if there
are qualitatively new types of signals in possible supersymmetric
extensions of the Standard Model (SM).

Since there is a well tested correspondence between
five-dimensional (5D) anti-de Sitter (AdS) space and
four-dimensional conformal field theories (CFT's)~\cite{Maldacena}
one would expect 5D Scherk-Schwarz breaking \cite{Scherk:1978ta}
of SUSY to provide a rough qualitative guide. In a SUSY 5D theory
with a finite extra dimension a SM fermion corresponds to two
Kaluza-Klein (KK) towers, one for the fermion modes and one for
the bosonic superpartner modes. Recall that Scherk-Schwarz
breaking allows us to shift the spectrum of the bosonic partners
of the quarks and leptons up in energy so that the superpartners
are no longer degenerate.  In a SUSY CFT we have two continua for
the bosons and fermions rather than two KK towers. We can break
the conformal symmetry by introducing a mass gap below the two
continua in such a way that there are still bosonic and fermionic
zero-modes (i.e. massless particles). The zero-modes should be
identified  as an $\mathcal{N}=1$ four-dimensional (4D)
supermultiplet. Now if we also break this residual SUSY by means
of a soft breaking mass,  we can lift the zero-mode of the boson.
It is not too hard to imagine that we can even push the bosonic
zero-mode all the way into the continuum, so that when we look for
a superpartner of the fermion, all we can find is a bosonic
continuum, that is an unparticle, or in this case a
``sunparticle.''

We will show by explicit computation, using the AdS/CFT
correspondence, that the scenario above can actually occur.  We
also find that in certain regions of parameter space, there are
many discrete modes below the continua, and that in the simplest
case the bosonic and fermionic gaps actually remain equal after
SUSY breaking.  This implies that new types of search techniques
are required at the LHC since even standard decay chains (e.g., the
gluino chain) can be drastically modified.

After reviewing the AdS/CFT correspondence, we study the effective
(holographic boundary) action corresponding to a chiral
supermultiplet, and establish the relation between the bosonic and
fermionic continua as well as the discrete levels.  We then
introduce SUSY breaking on the boundary of the AdS space and trace
through it consequences on the spectra.  We briefly comment on the
case of a vector supermultiplet, and describe some
phenomenological consequences.

% ----------------------------------------------------------------------------
\section{AdS/CFT Correspondence}

Using the AdS/CFT correspondence, we can study supersymmetric
unparticles in the context of RS2  scenarios \cite{RS2} without an IR brane.
Consider the 5D
AdS metric written in Poincar\'e coordinates:
\beq d s^2 = \left( \frac{R}{z} \right)^2 \left( \eta_{\mu \nu} d
x^\mu d x^\nu - d z^2 \right)\,.
\eeq
In the standard RS2 scenario, the space is cutoff below
$ z_{UV} = \epsilon$ (aka the UV-brane)  We can interpret $1/z$ as the renormalization scale so
that small and large $z$ correspond to the UV and IR
respectively of the corresponding dual 4D theory. Fields localized
in the UV-brane, which are not charged under bulk gauge symmetries,
act much like a hidden sector probing the CFT. This last observation has developed into
a technique called holography which we will summarize briefly.

 The procedure is the following: first one integrates over the bulk
fields, $\Phi$, constrained by a  UV-boundary condition,
$\Phi(x,z_{UV})=\Phi^0(x)$. This is done by using the equations of
motion (EOM) for the fields, thus obtaining an effective 4D
non-local action for the UV boundary fields $\Phi^0$. Since the theory is weakly coupled we
are allowed to use this classical (tree-level) approximation. Now there is
a correspondence between the
partition function corresponding to the 4D effective boundary
action and the generating functional obtained by integrating out a
4D strongly coupled CFT,
\begin{equation}
\mathcal{Z}[\Phi^0]=
e^{i[S_{eff}(\Phi^0)+S_{UV}(\Phi^0)]}=\int d\Phi_{CFT}\,
e^{i[S_{SCFT}+\Phi^0\mathcal{O}]}\label{partition}
\end{equation}
where on the right hand side of Eq.~(\ref{partition}), $\Phi^0$
plays the role of an external field that couples to the strongly
coupled  conformal sector through the CFT operators $\mathcal{O}$.
{}From this expression we see that $\Phi^0$ acts like a source for
the  CFT operator $\mathcal{O}$.
Thus for a CFT operator  ${\mathcal O}$ with a particular scaling dimension we can describe its
correlation functions using a 5D AdS action or a non-local 4D action for an ``unparticle \cite{Georgi,Falkowski,Cacciapaglia,Friedland:2009iy}.

Let us now add SUSY to this setup. We know
that in 5D the lowest number of supersymmetric charges we can have
is 8. Furthermore higher-dimensional supersymmetric theories
contain 4D supersymmetry, so it is  possible to
write them down using 4D ${\mathcal N} = 1$ superfields \cite{MartiPomarol}. Concentrating on the matter
fields, we decompose a 5D ${\mathcal N} =1$
hypermultiplet $\Psi$ into two 4D ${\mathcal N} =1$ chiral
superfields $\Phi=\{\phi, \chi, F\}$ and $\Phi_c=\{\phi_c, \psi,
F_c\}$, where the two Weyl fermions $\chi$ and $\psi$ form a Dirac
fermion. In other words  a 5D ${\mathcal N} =1$ theory corresponds to a 4D ${\mathcal N} =2$ theory. The bulk 5D
AdS action for such a hypermultiplet takes the
form~\cite{MartiPomarol}, \beq S & =& \int d^4 x\, d z \left\{
\int d^4 \theta\, \left( \frac{R}{z} \right)^3
\left[ \Phi^\ast\, \Phi + \Phi_c\, \Phi_c^\ast\right] + \right.\nonumber\\
& & \left.+ \int d^2 \theta\, \left( \frac{R}{z} \right)^3  \left[
\frac{1}{2}\; \Phi_c\, \partial_z \Phi - \frac{1}{2}\; \partial_z
\Phi_c\, \Phi + m \frac{R}{z}\; \Phi_c\, \Phi \right] + h.c.
\right\}\,,\label{5Daction} \eeq which is explicitly hermitian
without boundary terms and where $m$ is a $z$-independent bulk mass
term. We identify the left-handed fields with $\Phi$ and the right
handed fields with $\Phi_c$.

The supersymmetric case when $m R = c $ is a constant has been
analyzed in Ref.~\cite{Cacciapaglia:2008bi} where it was shown
that this theory has a simple correspondence to a 4D CFT. In the
case that the right-handed field is the source for a left-handed CFT chiral operator
$\mathcal{\bf O}_{L}=(\mathcal{O}_{L},\Theta_L,F_L)$
it was shown that for values of $c < 1/2$, the scaling dimension
of the scalar component $\mathcal{O}_{L}$ is
\beq d_s=\frac{3}{2}-c \eeq
and the chiral fermion operator
\cite{Cacciapaglia:2008bi,Contino:2004vy} $\Theta_L$ has a scaling
dimension \beq d_f=2-c\,. \eeq In fact the scaling dimensions of
$\mathcal {\bf O}_L$, independently of the chirality, obey the
relationships $d_{s}=d_f-1/2=d_F-1$ as a consequence of SUSY. Note
that, with this sign convention, the composite field is more
and more elementary as $c$ is increased toward $1/2$, and in fact
saturates the Unitarity bound \cite{Mack} $d_s\ge 1$, $d_f\ge 3/2$
at $c=1/2$.  As is the case in Seiberg duality, after reaching the
Unitarity bound the fields become free fields, so  for
 $c>1/2$, the CFT operator is a free
superfield, with canonical scaling
dimension~\cite{Cacciapaglia:2008bi}. For right-handed CFT
operators we have similar expressions, but with $c\rightarrow -c$.

In order to obtain a phenomenologically viable theory, we need to
generate a mass gap in the spectrum of particles charged under the
SM gauge group. This can be accomplished by
introducing a bulk mass of the form $ m(z) R = c + \mu z$ as was
studied in the non-SUSY case in Ref.~\cite{Cacciapaglia}.
We would like to comment at this point that such a
$z$-dependent mass term violates the 5D local Lorentz invariance
in the bulk as well as half of the SUSY. However, 4D Lorentz
invariance and ${\mathcal N}=1$ SUSY are manifestly preserved.
One can imagine that this $z$-dependent mass term comes from a
$z$-dependent vacuum expectation value (VEV) of some bulk field.
(For instance, a dilaton VEV can be transformed into such a term after field rescaling \cite{Falkowski}.)
The 5D  ${\mathcal N}=1$ SUSY and local Lorentz invariance
can be nonlinearly realized with inclusion of a Goldstone
supermultiplet~\cite{Bagger}.

{}From the 5D action, Eq.~(\ref{5Daction}), we derive the first order
coupled EOM for fermions,
\begin{eqnarray}
 - i\bar \sigma ^\mu  \partial _\mu \chi  - \partial _z \bar \psi  + (m(z)R + 2)\frac{1}{z} \bar \psi  &=& 0 ,\label{fermion1}\\
- i\sigma ^\mu  \partial _\mu  \bar \psi  + \partial _z \chi  +
(m(z)R - 2)\frac{1}{z}\chi  &=& 0\label{fermion2}.
\end{eqnarray}
Furthermore, we can easily obtain the expressions for the $F$-terms,
\begin{eqnarray}F_c^ * &=&  {-\partial _z \phi + \left(\frac{3}{2} - m(z)R\right)\frac{1}{z}} \phi \label{fc} ,\\
F &=&  {\partial _z \phi_c^* -\left(\frac{3}{2} +m(z)R\right)\frac{1}{z}}
\phi^*_c ,
\end{eqnarray}
and use them to find second order EOM for the scalars:
\beq
\partial_\mu \partial^\mu \phi - \partial_z^2 \phi + \frac{3}{z}\; \partial_z \phi + \left( m(z)^2 R^2+ m(z)R - \frac{15}{4} \right) \frac{1}{z^2}\; \phi  -\left(\partial_z m(z)\right) \frac{R}{z}\phi=
0\,.\label{scalar} \eeq where the EOM for $\phi_{c}$ is the same
as for $\phi$ but with $m(z)\rightarrow -m(z)$. Note that the
Breitenlohner--Freedman bound \cite{Breitenlohner} on the scalar mass in AdS
\beq
m(z)^2 R^2+ m(z)R - \frac{15}{4} >-4 ~,
\eeq
is automatically satisfied.

As usual, we can decompose the fields as products of a
4D fields times profiles in the extra dimension and solve for the
profiles. To do this it is convenient to write the
component fields of the hypermultiplet (in 4D momentum space) as,
\begin{eqnarray}
 \chi (p,z) & = &  \chi _{\rm{4}} (p)\left(\frac{z}{z_{UV}}\right)^2 f_L(p,z),  \qquad  \phi (p,z) =  \phi _{\rm{4}} (p)\left(\frac{z}{z_{UV}}\right)^{3/2} f_L(p,z), \label{decom1}\\
 \psi (p,z) & = &  \psi _{\rm{4}} (p)\left(\frac{z}{z_{UV}}\right)^2 f_R(p,z), \qquad
\phi _c (p,z) =  \phi _{{\rm{c4}}}
(p)\left(\frac{z}{z_{UV}}\right)^{3/2} f_R(p,z), \label{decom2}
 \end{eqnarray}
where the relationships between scalar and fermion profiles are
provided by SUSY, and $p\equiv \sqrt{p^2}$. Substituting Eqs.(\ref{decom1}--\ref{decom2}) into
Eqs.(\ref{fermion1}--\ref{fermion2}--\ref{scalar}), we find that
the solutions can then be expressed as linear combinations of the
Whittaker functions of the first kind and the second kind~\cite{FalkowskiWhitt},
\begin{eqnarray}
 f_L(p,z) &=& a \,M( \kappa,\frac{1}{2} + c,2\sqrt {\mu^2  - p^2 } z)
+ b \, W( \kappa,\frac{1}{2} + c,2\sqrt {\mu^2  - p^2 } z)\label{fL}~,
 \\
f_R(p,z) &=& a \, \frac{2(1 + 2c)\sqrt {\mu^2  - p^2 }}{p} M( \kappa, - \frac{1}{2} + c,2\sqrt
{\mu^2 - p^2 } z) \nonumber \\ &&+\, b \, \frac{p}{(\mu +\sqrt
{\mu^2 - p^2 } )} W( \kappa, -
\frac{1}{2} + c,2\sqrt {\mu^2 - p^2 }z)\label{fr}~,\\
\kappa&\equiv&- \frac{{c\,\mu}}{{\sqrt {\mu^2  - p^2 } }}~,
\end{eqnarray}
where $M(\kappa,m,\zeta)$ is the Whittaker function of the first
kind which in its series form is given by,
\beq M(\kappa,m,\zeta)&=&\zeta^{m+1/2}
e^{-\zeta/2}\sum_{n=0}^\infty \frac{ \Gamma(m-\kappa+1/2+n)
\Gamma(2m+1) } {n! \Gamma(m-\kappa+1/2)\Gamma(2m+1+n)} \,\zeta^n~,
\label{Mdef}\eeq
and $W(\kappa,m,\zeta)$ is the Whittaker function of the second
kind which can be expressed as,
\beq W(\kappa,m,\zeta)&=&\frac{ \Gamma(-2m) }
{\Gamma(1/2-m-\kappa)} M_{\kappa,m}(\zeta)+\frac{ \Gamma(2m) }
{\Gamma(1/2+m-\kappa)} M_{\kappa,-m}(\zeta).\label{Wdef} \eeq
{}From the fermionic EOM, Eq.~(\ref{fermion1}-\ref{fermion2}), we
conclude that $f_L(p,z)$ and $f_R(p,z)$ are related to each other
by,
\begin{eqnarray}
 \left( {\partial _z  + \frac{1}{z}(c + \mu z)} \right)f_L (p,z) &=& p f_R (p,z)\label{1storder1}, \\
 \left( {\partial _z  - \frac{1}{z}(c + \mu z)} \right)f_R (p,z) &=& -p f_L
 (p,z)\label{1storder2}.
 \end{eqnarray}
It is simple to solve for the zero mode profiles by looking at
Eqs.(\ref{1storder1}--\ref{1storder2})  in the case $p^2=0$. Then
the zero modes are given by $ f_L(0,z) \sim e^{ - \mu z} z^{ - c}$
and $f_R(0,z) \sim e^{\mu z} z^{c}$. Thus, in order to have a
normalizable mode in the sense that the wave function vanishes
when $z\rightarrow +\infty$, we notice that for positive values of
$\mu$ only the left-handed zero mode is normalizable, while  the
right-handed zero mode is normalizable for negative $\mu$.

As previously mentioned, Eqs.(\ref{1storder1}--\ref{1storder2})
relate the profiles $f_L (p,z)$ and $f_R(p,z)$ for non-vanishing
momenta $p$. The ratio of the coefficients $a/b$ is fixed by the
asymptotic behavior as  $z\rightarrow +\infty$.

It is illuminating to rewrite the equations of motion for the
right-handed and left-handed fields in a form analogous to
ordinary quantum mechanics. Let us take a look at the fermion
EOMs. Rescaling the fields we can write the equations for the
profiles in the form,
\begin{eqnarray}
\frac{\partial^2}{\partial z^2}f_R+\left(p^2-\mu^2-2\frac{\mu
c}{z}-\frac{c(c-1)}{z^2}\right)f_R&=&0\label{2ndorder1},\\
\frac{\partial^2}{\partial z^2}f_L+\left(p^2-\mu^2-2\frac{\mu
c}{z}-\frac{c(c+1)}{z^2}\right)f_L&=&0 \label{2ndorder2}.
\end{eqnarray}
This can be compared with the radial Schr\"{o}dinger equation for the
Hydrogen atom with a reduced mass $M$: \beq
\frac{\partial^2}{\partial r^2}u+\left(2 M E+2\frac{M
\alpha}{r}-\frac{\ell(\ell+1)}{r^2}\right)u&=&0 \eeq where $E$ is
the negative binding energy. We can therefore identify the
following expressions for the effective potentials for $f_R$ and
$f_L$ respectively,
\begin{equation}
V_R(z)=\frac{c(c-1)}{z^2}+2\frac{c\mu}{z},\qquad
V_L(z)=\frac{c(c+1)}{z^2}+2\frac{c\mu}{z}.\label{potential}
\end{equation}
Because of SUSY these potentials govern the behavior of the scalar
components as well as fermions, though the rescaling necessary to
obtain them is different for scalars. We notice that for $V_R(z)$
in the case of negative values of $c$ and positive values of
$\mu$, the potential is exactly the one corresponding to the
Hydrogen atom with the identifications of angular momentum
$\alpha=\ell=|c|$. Similarly, for $V_L(z)$ in the case of positive
$c$ and negative $\mu$, we find again the Hydrogen potential with
a reduced mass $-\mu$ and the identifications and $\alpha=\ell=c$.
Notice that in the case of negative $\mu$ for the right-handed
field, there is a corresponding zero mode associated with it. Thus
in these regions of parameter space there is an analogous
structure to the energy levels of the Hydrogen atom, that is an
infinite tower of KK states below the mass gap (i.e. with $p^2<
\mu^2$). This last statement is verified analytically and in
numerical examples in the next section.

We will concentrate in the case $\mu>0$ where the left-handed chiral
supermultiplet acquires a normalizable zero mode for momentum
smaller than the mass gap. In analyzing the spectra we will pay
special attention to the range $-1/2<c<1/2$ that corresponds to a
scaling dimension for the scalar, $1<d_s<2$. One can use two
different approaches to obtain the spectrum: by applying the boundary
conditions to the bulk solutions in order to  find the
extra-dimensional wavefunctions of the modes directly,
or by calculating the holographic effective action and
probing the CFT operators with a right-handed superfield source.
We leave the first approach for the Appendix and concentrate on
the second in the next section.

%%%%%%%%%%%%%%%%%%%%%%%%%%%%%%%%%%
\section{Holographic Boundary Action}

As mentioned above, one method for obtaining the spectrum is by
calculating the holographic boundary action. In this way, we can
probe the CFT by sourcing it with a boundary superfield. First
we integrate out the bulk field constrained on the UV
brane by the boundary condition $\Phi_c(z_{UV})=\Phi_{c}^{0}$.
Here $\Phi_{c}^{0}$ plays the role of a source for the CFT  in the
holographic interpretation while the field $\Phi$ is allowed to
vary freely at the UV brane ~\cite{Cacciapaglia:2008bi}. Thus, we
add the following UV boundary superpotential term which will
cancel the UV variation coming from $\Phi$,
\begin{equation}
S_{UV}=-\int
d^4x\frac{1}{2}\left(\frac{R}{z_{UV}}\right)\left(\int d^2\theta
\Phi(z_{UV})\Phi_{c}^{0}+h.c\right).
\end{equation}
Requiring the variation of the action to vanish on the UV brane,
the boundary conditions read,
\begin{equation}
F_c(z_{UV} ) = F_{c}^{0}   , \qquad \psi (z_{UV} ) = \psi^0 , \qquad
\phi_c (z_{UV} ) = \phi _{c}^{0}.
\end{equation}

After integrating the bulk, we get the supersymmetric holographic
action:
\begin{eqnarray}
S_{holo}  =  - \int {d^4 } x[\phi _{c}^{0*}  \Sigma_{\phi_c} \phi
_{c}^{0}  + F_{c}^{0*}  \Sigma_{F_c} F_{c0}  + \psi _{0}^ *
\Sigma_{\psi} \psi _{0} ]\label{holoaction}
\end{eqnarray}
where
\begin{equation}
\Sigma_{\phi_c}=\left(\frac{R}{z_{UV}}\right)^3p\frac{f_L}{f_R},\quad
\Sigma_{\psi}=\left(\frac{R}{z_{UV}}\right)^4\frac{p_\mu\sigma^\mu}{p}\frac{f_L}{f_R},\quad
\Sigma_{F_c}=\left(\frac{R}{z_{UV}}\right)^3\frac{1}{p}\frac{f_L}{f_R}.
\end{equation}
{}From the CFT point of view,  the right-handed superfield
$\Phi_{c}^{0}$ is a source for a left-handed chiral superfield as
a CFT operator that couples to. Furthermore, since $F_c$ is the
source for the scalar component, as was shown
in~\cite{Cacciapaglia:2008bi}, the propagator for the scalar CFT
operator is given by, \beq \Delta_{s}(p)\propto-\Sigma_{F_c}(p).
\eeq Moreover, as a consequence of SUSY we also have that the
fermionic and $F$-component correlators of the CFT operator are
related to $\Delta_{s}$ by $\Delta_{f}=p_\mu\sigma^\mu\Delta_{s}$
and $\Delta_{F}=p^2\Delta_{s}$.

We adopt outgoing wave boundary conditions and therefore drop the
Whittaker function of the first kind, $M(\kappa,m,z)$, in
Eqs.(\ref{fL}--\ref{fr})~\cite{Giddings:2000mu}. This is
equivalent to Wick rotating to Euclidean momenta and keeping
only the solution that decays exponentially for large Euclidean
momenta~\cite{Falkowski}.
Now we concentrate on the study of the scalar propagator in the
conformal limit $\epsilon\rightarrow 0$. For that purpose we
expand $\Sigma_{F_c}$ for small values of $z_{UV}=\epsilon$,
\begin{eqnarray}
\Sigma_{F_c} =  \frac{\epsilon(\mu +\sqrt {\mu^2  - p^2 })}{p^2
}\cdot \frac{{W\left( { - \frac{{c\mu}}{{\sqrt {  \mu^2 - p^2 }
}},\frac{1}{2} + c,2\sqrt { \mu^2 - p^2  } \epsilon}
\right)}}{{W\left( { - \frac{{c\mu}}{{\sqrt {  \mu^2 - p^2 }
}},\frac{1}{2} - c,2\sqrt {  \mu^2 - p^2 } \epsilon} \right)}},
\end{eqnarray}
and concentrate in the region $-1/2<c<1/2$. We find that after
properly rescaling the correlator by a power of $\epsilon$ (to obtain the correct
dimension of the correlator),
\begin{equation}
\Delta_{s}^{-1}\approx { \frac{{p^2 }}{{1 - 2c}}\epsilon^{1-2c} - \frac{{2^{ - 1 + 2c} p^2 (\mu^2  - p^2 )^{ - 1/2 + c}
\Gamma (1 - 2c)\Gamma (c + \frac{{c\mu}}{{\sqrt {\mu^2  - p^2 }
}})}}{{\Gamma (2c)\Gamma (1 - c + \frac{{c\mu}}{{\sqrt {\mu^2  -
p^2 } }})}}}. \label{delta}\end{equation}

Notice that in the case $-1/2<c<1/2$, since $\epsilon ^{ 1-2c}
\rightarrow 0$ in the conformal limit, the first term in
Eq.~(\ref{delta}) can be ignored. In this limit, we
notice that there is always a massless mode associated with the
CFT operator and therefore the CFT is chiral. We see from this
last expression that effectively there is pole located at $p^2=0$
which corresponds to the zero mode found in Eq.~(\ref{1storder1}).
Furthermore, when $p^2>\mu^2$, we find a branch-cut and thus the
beginning of the continuum. In the case $-1/2<c<0$, besides the
massless pole, there are momenta in the range $p^2 < \mu^2$ where
the function $\Gamma (1 - c +c\mu/(\sqrt {\mu^2  - p^2}))$ can
also become  large  and it is possible for the first and second
terms to be comparable and cancel each other. This leads
to exactly the condition found in the Appendix, Eq.~(\ref{poles}),  by
solving for the extra-dimensional wavefunctions using
Eqs.~(\ref{2ndorder1}) and (\ref{2ndorder2}). The corresponding
series of poles with $p^2\sim\mu^2$ can be related to the
Hydrogen-like solutions expected from Eq.~(\ref{potential}). This
series of poles remains in the conformal limit. Again, for
$p^2>\mu^2$ we find a branch-cut that signals the beginning of the
continuum.

In the case of the continuum spectrum above the mass gap,
$p^2>\mu^2$, the spectral density function is,
\begin{equation}
\rho(p^2)= \frac{1}{R_0} {\rm Im}\left[\Sigma_{F_{\phi_c}}\right]\, .
\end{equation}
and we have normalized it\footnote{We are forced to normalize the spectral
function in this way since in the range $-1/2<c<1/2$, the integral
$\int^{1/\epsilon}_{\mu}\rho(p^2)pdp$ diverges as a power of the cut-off.}
with respect to the residues of the corresponding Standard Model
(SM) zero-mode fields with $R_{0}$ being the residue coming from the zero-mode pole. The
expression for the zero mode residue is very simple to obtain
since we already know that the pole comes from the pre-factor in
Eq.~(\ref{fr}). Thus we have,
\begin{equation}
R_{0}=2\mu\epsilon \frac{
W(-c,1/2+c,2\mu\epsilon)}{W(-c,-1/2+c,2\mu\epsilon)}\label{norm}
\end{equation}
%
%For completeness, in the region $-1/2<c<0$, we find that in the
%limit $\epsilon\rightarrow 0$, the residues of the other KK poles
%are
%
%\begin{equation}
%R_n=\frac{2^{1-2c}c^2(c\mu)^{-1-2c}\Gamma(n+1-2c)}{\Gamma(1-2c)n!(1+n)(1-c+n)^{-1-2c}}\label{residue}
%\end{equation}
%

\begin{figure}[htbp]
      \centering
        \includegraphics[scale=0.7]{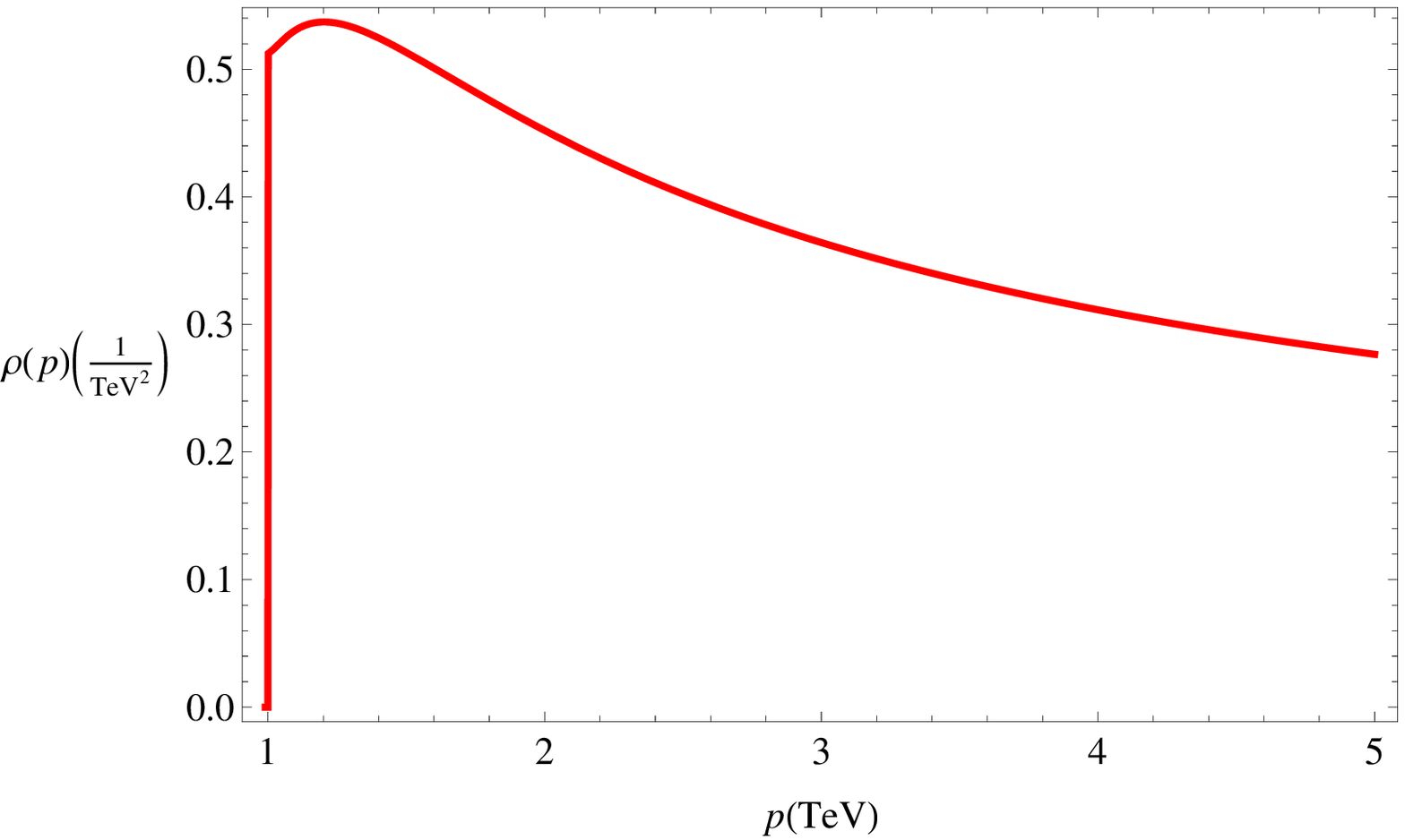}
    \centering
        \includegraphics[scale=0.68]{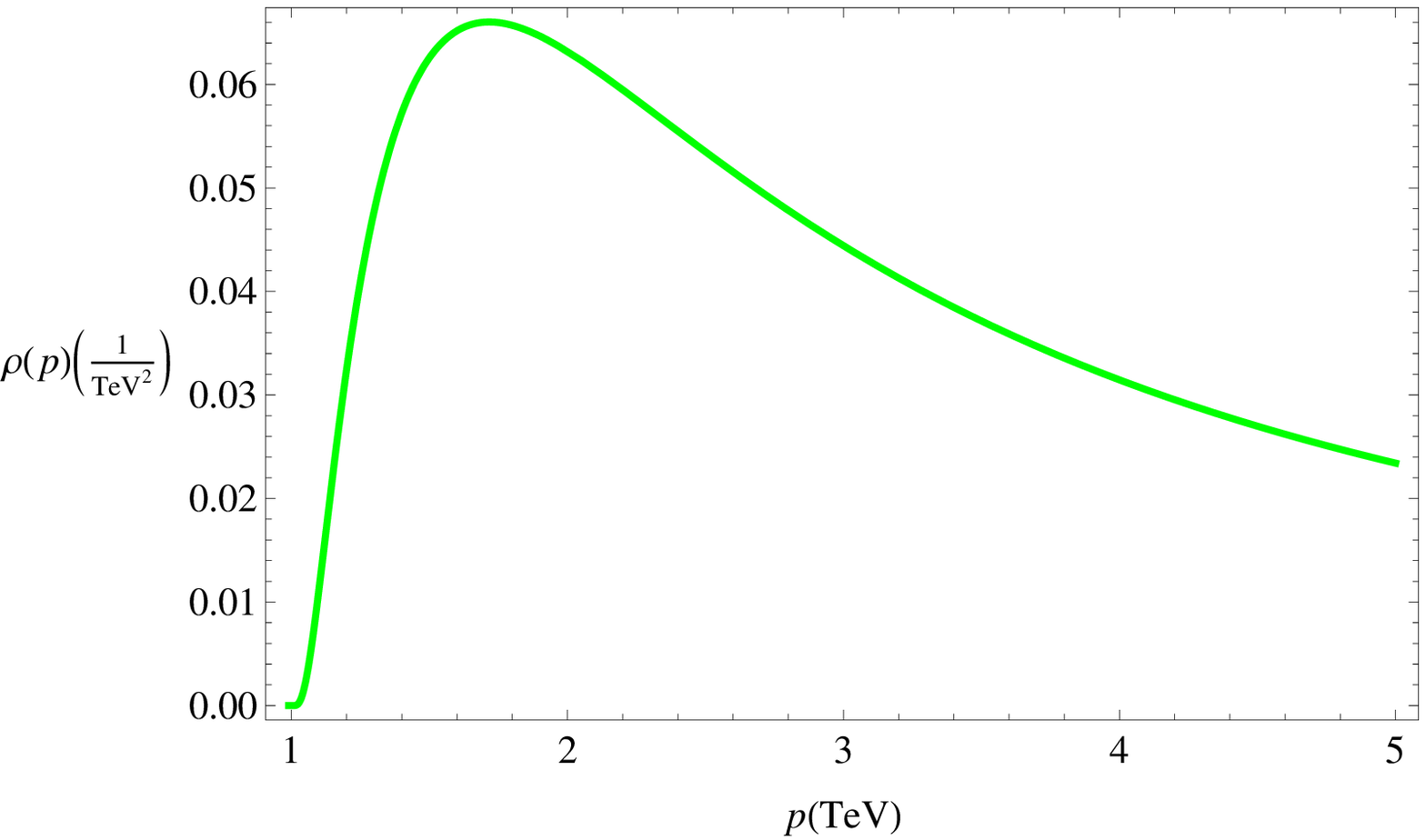}
\caption{Continuum spectral density function with
$\epsilon=10^{-19}\;\rm{GeV}^{-1}$ and $\mu=1$ TeV, as a function
of momenta $p$ in TeV. In the first example (red curve) $c=-0.3$
and in the second example (green curve) $c=0.3$.}\label{continuum}
\end{figure}

In Fig.~\ref{continuum} we plot the spectral density as a function
of momentum for two values of $c$. The spectral density peaks at
higher momenta for higher values of $c$.

In the region $c>1/2$, for small  $\epsilon$, the correlator
reduces after proper rescaling to,
\begin{equation}
\Delta_{s}^{-1}\approx
p^2\left(\frac{\Gamma(2c-1)}{\Gamma(2c)}+\epsilon^{2c-1}
\frac{2^{-1+2c}(\mu^2-p^2)^{-1/2+c}\,\Gamma(1-2c)
\Gamma(c+\frac{c\mu}{\sqrt{\mu^2-p^2}})}{\Gamma(2c)\Gamma(1-c+
\frac{c\mu}{\sqrt{\mu^2-p^2}})}\right)\, .
\end{equation}
and in the limit $\epsilon\rightarrow 0$ we get,
\begin{equation}
\Delta_{s,c}\approx \frac{2c-1}{p^2}\, .
\end{equation}
Notice that there is still a massless pole, corresponding to a
free field in the CFT. On the other hand, when $c<-1/2$, after
proper normalization, we find,
\begin{eqnarray}
\Delta_{s}&\approx&
-\frac{2(p^2-\mu(\mu+\sqrt{\mu^2-p^2}))\Gamma(-1-2c)\Gamma(1-c+
\frac{c\mu}{\sqrt{\mu^2-p^2}})}{p^2\,\epsilon^{-1-2c}\Gamma(1-2c)\Gamma(-c+\frac{c\mu}{\sqrt{\mu^2-p^2}})}\nonumber\\
&+& \frac{2^{1-2c}(\mu^2-p^2)^{1/2-c}\Gamma(2c)\Gamma(1-c+
\frac{c\mu}{\sqrt{\mu^2-p^2}})}{p^2\Gamma(1-2c)\Gamma(c+\frac{c\mu}{\sqrt{\mu^2-p^2}})}
\end{eqnarray}
which in the limit $p^2\rightarrow 0$ reduces to,
\begin{equation}
\Delta_{s}\approx\frac{1}{\epsilon^{-1-2c}(1+2c)}+
\frac{2^{1-2c}\mu^{1-2c}}{p^2\Gamma(1-2c)}\, .
\end{equation}
Notice there is a UV sensitivity that can be cancelled by adding a
proper SUSY term in the boundary
action~\cite{Cacciapaglia:2008bi}. The massless pole always
remains in the spectrum, even in the CFT limit $\epsilon$.
Furthermore, its existence depends completely on $\mu$, since when
$\mu$ vanishes the massless pole disappears from the spectrum.

%%%%%%%%%%%%%%%%%%%%%%%%%%%%%%%
\section{SUSY breaking}

Now we introduce SUSY breaking on the boundary
by means of a scalar mass term,
\begin{equation}
\delta S = \frac{1}{2} \int {d^4 x}\left( \frac{R}{z_{UV}} \right
)^3 \int {dz\left( {m^2 z_{UV} \cdot \phi ^ *  \phi  + h.c.}
\right)} \delta (z - z_{UV})\, .\label{boundarysusybreaking}
\end{equation}
The mass  $m$ has
dimension one, since a scalar field has dimensions
$d[\phi]=3/2$. This new term has the effect of modifying the
boundary conditions, which now take the form,
\begin{equation}
F_c(z_{UV} ) = F_{c}^{0}  + m^2 z_{UV}\, \phi^ * (z_{UV}) , \qquad
\psi (z_{UV} ) = \psi^0 , \qquad \phi_c (z_{UV} ) = \phi _{c}^{0}
\end{equation}

Using Eq.~(\ref{fc}), Eq.~(\ref{1storder1}) and
Eq.~(\ref{boundarysusybreaking}), we find that $\Sigma_{F_c}$ now
corresponds to,
\begin{equation}
\Sigma_{F_{c}}(p^2)=\left(\frac{R}{z_{UV}}\right)^3\frac{f_L}{pf_{R}-m^2\,z_{UV}f_L}\label{susybreaking1}\,.
\end{equation}

As was done in the previous section, we study the new scalar
propagator modified by SUSY breaking in the CFT limit
$\epsilon\rightarrow 0$. In this case, in the region of bulk mass,
$-1/2<c<1/2$, after proper normalization of the correlator, we
find,
\begin{equation}
\Delta_{s}^{-1}(p^2) \approx  m^2 \epsilon ^{  1
-2c}-\frac{p^2}{2c-1}\epsilon^{1-2c} - \frac{2^{ - 1 + 2c} p^2
(\mu^2 - p^2 )^{ - 1/2 + c} \Gamma (1 - 2c)\Gamma (c +
\frac{c\mu}{\sqrt {\mu^2 - p^2 }})}{\Gamma (2c)\Gamma (1 - c +
\frac{c\mu}{\sqrt {\mu^2 - p^2 } })}
\label{susybreaking2}\end{equation}

\begin{figure}[htb]

        \centering
        \includegraphics[scale=0.7]{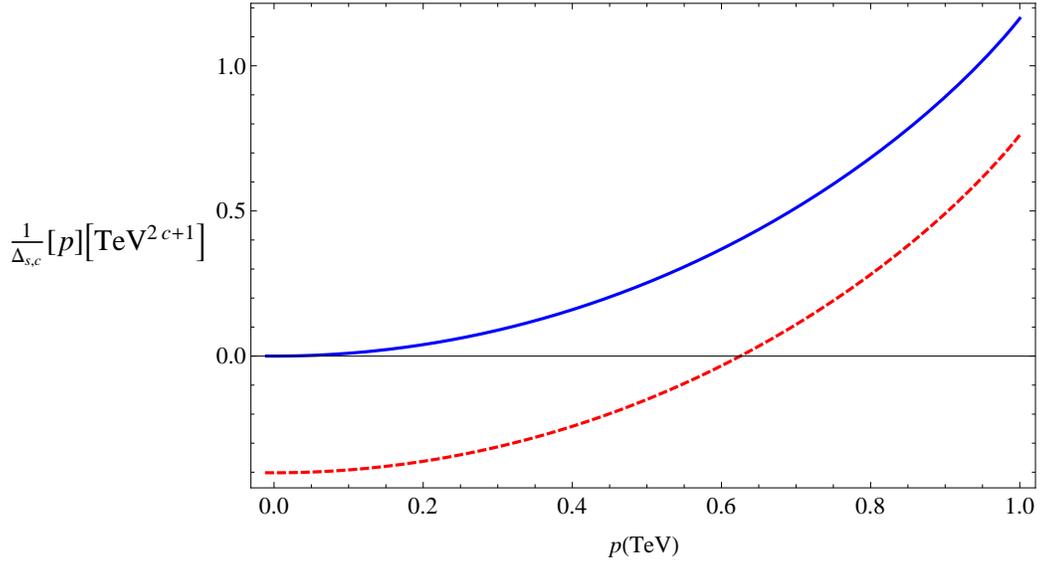}
        %\caption{Red represents the region where $\tilde{\nu}_{\tau}$ is the NLSP and green is the region where
        %$m_{\tilde{\chi}_{0}}>m_{\tilde{\nu}_{\tau}}$.
        %\label{collfig23}

\caption{Inverse correlator in the case of positive $c$ with,\
$m=0$ GeV (blue curve, solid) and $m=4\times 10^{7}$ GeV (red
curve, dashed), $\epsilon=10^{-19}\;\rm{GeV}^{-1}$, $\mu=1$ TeV
and $c=0.2$. Notice how the pole has shifted from $p^2=0$ at $m=0$
to $p^2\approx 600\; \rm{GeV}^{2}$ at $m=4\times 10^{7}$ GeV.
}\label{SUSYbreaking2}
\end{figure}

In the limit when $p^2\ll\mu^2$ we can easily solve for the new
displaced poles. In this case, we find that the pole which was
at zero momentum in the SUSY conserving action has been
displaced to a non-zero value  due to the SUSY breaking
mass term we added to the scalar action. The new pole location is
at
\begin{equation}
p^2_{pole}=\frac{2(2c-1)m^2(\mu\epsilon)^{1-2c}}{(-4^{c}+2^{1+2c}c)\Gamma(1-2c)}. \label{displacedpole}
\end{equation}
\begin{figure}[htb]

        \centering
        \includegraphics[scale=0.7]{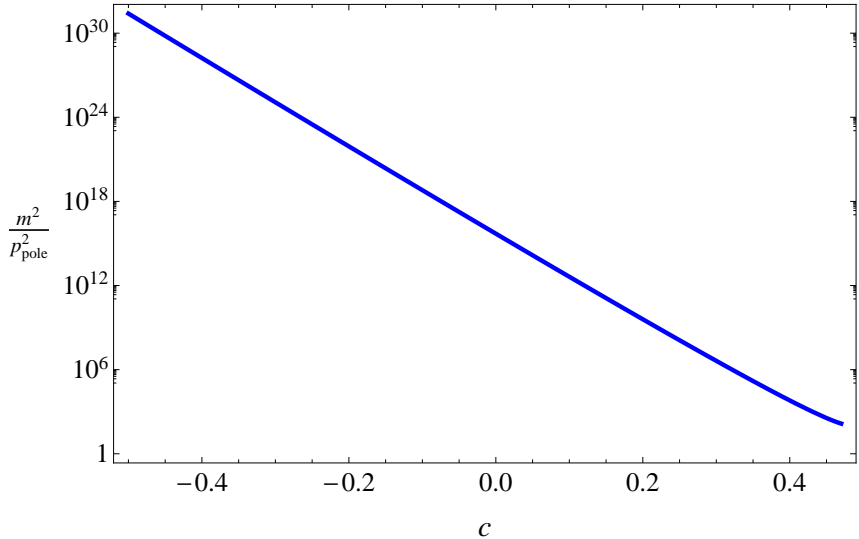}
        %\caption{Red represents the region where $\tilde{\nu}_{\tau}$ is the NLSP and green is the region where
        %$m_{\tilde{\chi}_{0}}>m_{\tilde{\nu}_{\tau}}$.
        %\label{collfig23}

\caption{Plot of $m^2/p^2_{pole}$ vs $c$. Notice that as $c$ gets
closer to $1/2$, for a given value of $p^2_{pole}$, the value of
$m^2$ decreases. }\label{logm2}
\end{figure}
We plot in Fig.~\ref{logm2} the dependence of the ratio
$m^2/p^2_{pole}$ as a function of the constant bulk mass $c$. An
important thing to notice is that as $c$ goes from $-1/2$ to
$1/2$, for a given value of $p^2_{pole}$, the necessary value of
$m^2$ decreases. This is easy to understand, since as $c$ becomes
negative, the zero-mode profile,
\begin{equation}
f_{L,0}\propto e^{-\mu z}z^{-c},
\end{equation}
is less localized near the UV brane and therefore is less affected
for a given SUSY breaking mass $m$. In Fig.~\ref{SUSYbreaking2} we
give a specific example of how the pole shifts for a given SUSY
breaking mass; the position of the pole is given by the zero of
the inverse correlator.

\begin{figure}[htb]

        \centering
        \includegraphics[scale=0.7]{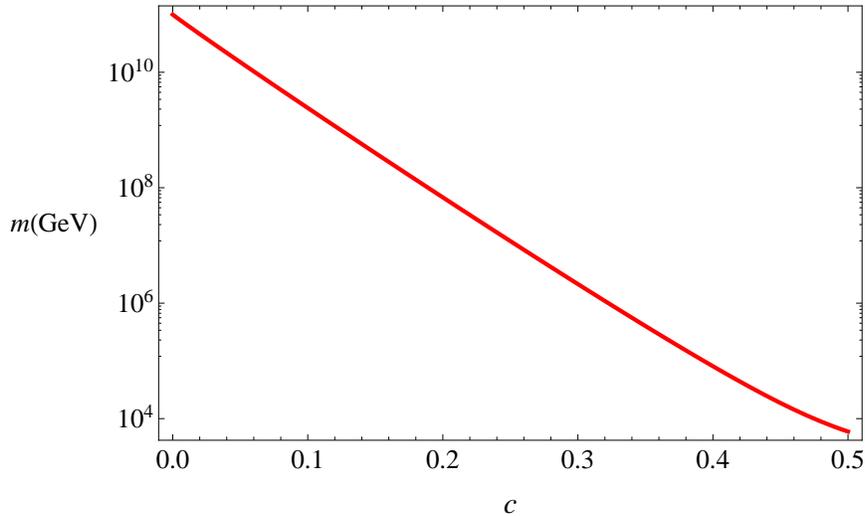}
        %\caption{Red represents the region where $\tilde{\nu}_{\tau}$ is the NLSP and green is the region where
        %$m_{\tilde{\chi}_{0}}>m_{\tilde{\nu}_{\tau}}$.
        %\label{collfig23}

\caption{SUSY-breaking mass vs bulk constant mass parameter $c$,
such that the zero mode pole exactly merges with the continuum. In
this example, $\mu=1$ TeV and $\epsilon=10^{-19}\;\rm{GeV}^{-1}$.
}\label{displacedpoles}
\end{figure}

For $c>0$, the zero mode is the only pole below the continuum. As
$m$ increases, the zero-mode pole moves closer into the continuum
with which it eventually merges. Thus, we see that one possibility
is a superpartner which only has a continuum spectrum, i.e.,
unparticle behavior. This has important consequences for
phenomenology which will be briefly discussed later. We can get an
approximate analytical expression for the displaced zero mode
poles in the vicinity of $p\approx \mu$. For that, we solve
Eqs.(\ref{2ndorder1}-\ref{2ndorder2}) with the constraint of
Eq.(\ref{1storder1}-\ref{1storder2}) for $p=\mu-\zeta$, with
$\zeta\ll\mu$. Replacing the solutions in Eq.(\ref{susybreaking1})
and analyzing the result in the CFT limit, $\epsilon\rightarrow
0$, we find that the pole, in the case of $0<c<1/2$, has shifted
to
\begin{equation}
p_{pole\approx\mu}=\frac{\mu}{2}-\frac{1}{2}\left(\frac{(1-2c)m^2}{\mu}-\frac{4^c
c^{2c}\epsilon^{-1+2c}\mu^{2c}\Gamma(2-2c)}{\Gamma(1+2c)}\right),
\end{equation}
so for
\begin{equation}
m^2 \gtrsim \frac{\mu^2}{1-2c} \left[ \frac{4^c c^{2c}\Gamma(2-2c) (\mu \epsilon)^{2c-1}}{\Gamma(1+2c)} -1\right]
\end{equation}
the zero mode merges into the continuum.
The value of the SUSY-breaking mass on the UV boundary where the pole just merges into the continuum
as a function of $c$ is plotted in
Fig.~\ref{displacedpoles}.

\begin{figure}[htbp]

        \centering
        \includegraphics[scale=0.9]{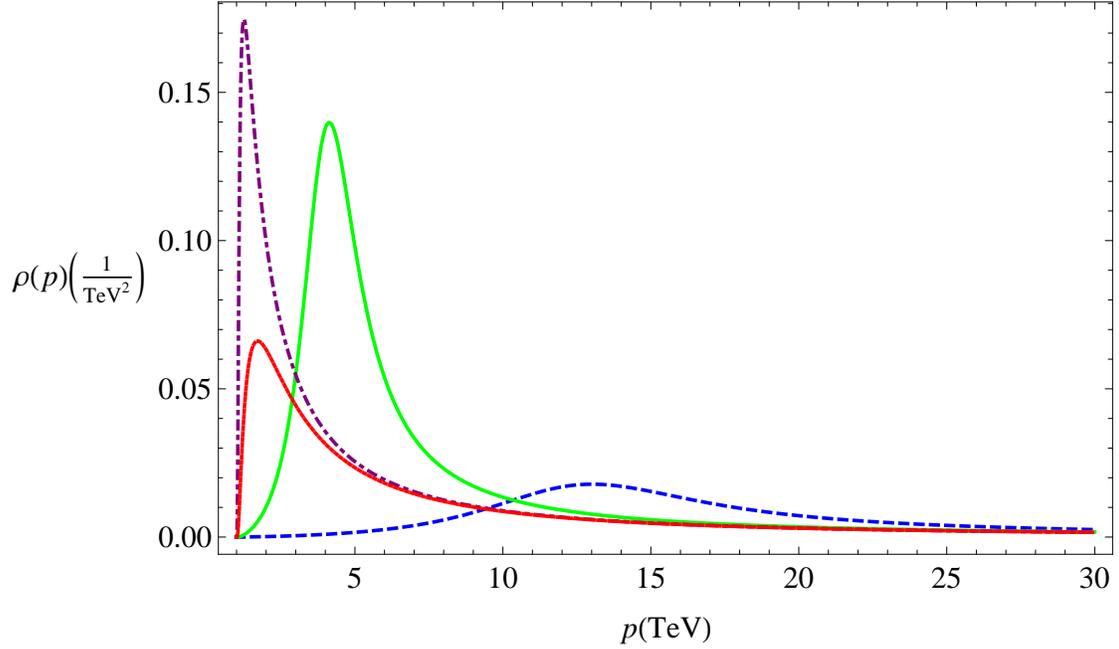}
        %\caption{Red represents the region where $\tilde{\nu}_{\tau}$ is the NLSP and green is the region where
        %$m_{\tilde{\chi}_{0}}>m_{\tilde{\nu}_{\tau}}$.
        %\label{collfig23}

\caption{Spectral function for four examples of boundary UV SUSY
breaking masses $m=2\times10^{7}$ GeV (blue curve, dashed),
$m=8\times 10^{6}$ GeV (green curve, solid), $m=2\times 10^{6}$
GeV (purple curve, dot-dashed) and $m= 10^{5}$ GeV (red curve,
dotted). The red and purple curves correspond to zero-mode poles
localized at $p\approx 50$ GeV (red curve) and $p\approx 950$ GeV
(purple curve), that haven't merged with the continuum. On the
other hand, the green and blue curves correspond to the cases
where the pole has merged into the continuum. In the examples,
$\epsilon=10^{-19}\;\rm{GeV}^{-1}$, $\mu=1$ TeV and $c=0.3$. We
can see how the continuum peaks to higher momenta as the SUSY
breaking mass $m$ increases, specially as the pole merges into the
continuum. }\label{continuum3}
\end{figure}

The shape of the continuum spectrum is also modified in the
presence of a SUSY breaking mass term. As mentioned previously, we
normalized with respect to the SM fields pole residues. We plot
the spectral density functions for several different SUSY-breaking
masses in Fig.~\ref{continuum3}. We see that  the peak shifts to
larger values of momenta with increasing values of $m$, in
particular after the pole merges with the continuum.

\begin{figure}[htb]

        \centering
        \includegraphics[scale=0.7]{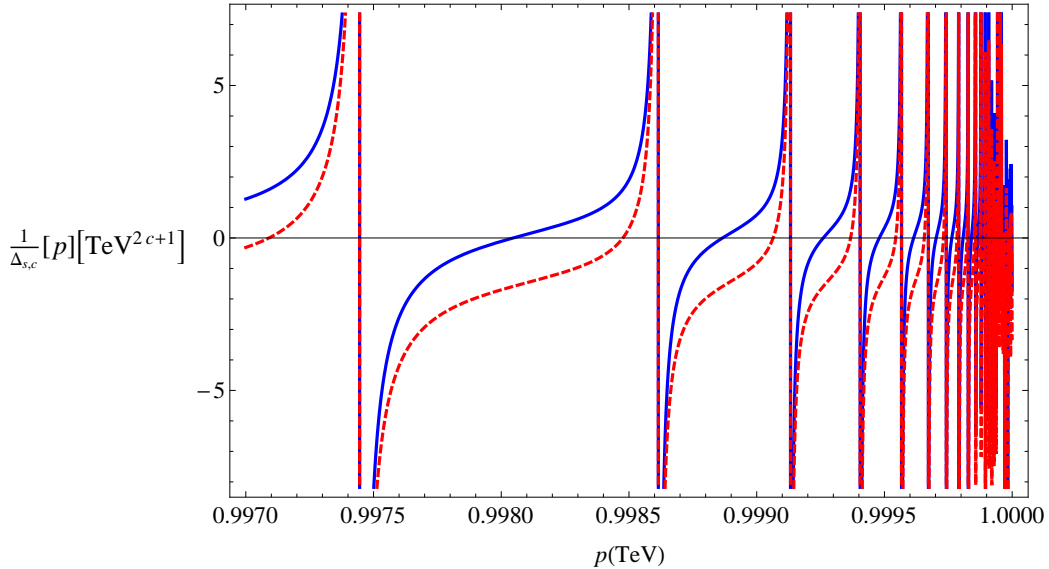}
        %\caption{Red represents the region where $\tilde{\nu}_{\tau}$ is the NLSP and green is the region where
        %$m_{\tilde{\chi}_{0}}>m_{\tilde{\nu}_{\tau}}$.
        %\label{collfig23}

\caption{Inverse correlator for two examples of boundary UV SUSY
breaking masses: $m=10^{13}$ GeV (blue curve) and $m=2\times
10^{14}$ GeV (red curve, dashed). In the examples,
$\epsilon=10^{-19}\;\rm{GeV}^{-1}$, $\mu=1$ TeV and $c=-0.2$. We
can see how the series of poles shift into the continuum with
increasing values of m. }\label{SUSYbreaking}
\end{figure}

For $c<0$, there are also resonant KK-like states below the
continuum besides the zero mode. Looking at
Eq.~(\ref{susybreaking2}), we notice that as $m^2$ increases, the
resonant KK-like states for $c<0$ will also move towards the
continuum. However, they never disappear into the continuum as the
last term in Eq.~(\ref{susybreaking2}) can become arbitrarily
large, when $c+(c\mu)/(\sqrt{\mu^2-p^2})$ gets close to a negative
integer, to compensate the first term independently of how large
$m$ is. It is difficult to obtain an analytical expression for the
displaced poles. However we can see the effects numerically as
shown in Fig.~\ref{SUSYbreaking}, where the poles are given by the
positions of the zeroes in the plot.

We can obtain an approximate analytical expression for the shift
of the KK-like resonances below the continuum for $c<0$ in the limit that the first term in
Eq.(\ref{susybreaking2}) dominates over the second term. In that
case we expect small modifications to the SUSY spectrum and we
find that the following relationship is satisfied,
\begin{equation}
1-c+\frac{c\mu}{\sqrt{\mu^2-p^2}}=-n+C~, \label{shiftedspectrum}
\end{equation}
where the constant $C$ is given by,
\begin{equation}
C=\frac{(-1)^{n+1}4^{-c}(1+n)c^{-2c}\Gamma(2c)}{(1-2c+n)\Gamma(-2c)\Gamma(2+n)^2(1+c+n)^{-1-2c}}\left(\frac{m^2}{\mu^2}\right)(\epsilon \mu)^{1-2c}~.
\end{equation}
Notice that for $m^2\rightarrow 0$, $C$ vanishes, recovering the
unperturbed solution.

Again, for completeness we analyze the case $c>1/2$. We find that
the correlator, in the limit $\epsilon\rightarrow 0$ reduces to,
\begin{equation} \Delta_{s}^{-1}(p^2)
\approx m^2-\frac{p^2}{2c-1}~.
\end{equation}
where  the pole has shifted as expected. For $c<-1/2$, after
proper normalization, we find,
\begin{eqnarray}
\Delta_{s}(p^2)&\approx&\frac{1}{
(1+2c)\epsilon^{-2c-1}}+\frac{2^{1-2c}\mu^{1-2c}}{p^{2}\,c(2+4c+(-2m^2+(1+2c)\mu^2)\epsilon^2)\Gamma(-1-2c)}
\label{shiftpole}
\end{eqnarray}
where once again we notice how the pole has been displaced by the
non-zero soft mass. In the exact CFT limit we notice a UV
sensitivity as we found previously for this region of parameter
space in the SUSY case.

%%%%%%%%%%%%%%%%%%%%%%%%%%%%%%%%%%%%%%%%%%%%%%%%%%%%%%%%%%%%%%%%%%%%%%%%%%%%%%%%%%%%
%%%%%%%%%%%%%%%%%%%%%%%%%%%%%%%%%%%%%%%%%%%%%%%%%%%%%%%%%%%%%%%%%%%%%%%%%%%%%%%%%%%%
\section{Gauge fields}

In this section we discuss the gauge fields. As was shown in
\cite{MartiPomarol}, a 4D ${\mathcal N}=2$ vector supermultiplet
can be decomposed into an ${\mathcal N}=1$ vector supermultiplet
$V=(A_{\mu},\lambda_{1},D)$ and a chiral ${\mathcal N}=1$
supermultiplet
$\sigma=((\Sigma+iA_5)/\sqrt{2},\lambda_{2},F_{\sigma})$. We
cannot proceed as before by introducing a bulk mass term for the
gauge fields since this would break gauge invariance. Thus, we add
a dilaton superfield interaction\footnote{We
could have proceeded in a similar fashion for the hypermultiplet
case and added a dilaton interaction in addition to the bulk
$z$-dependent mass. However, the dilaton VEV can be absorbed  by a
wavefunction redefinition and as shown below, this would only
provide a shift in the particular value of $\mu$ considered.} that softly
breaks the conformal symmetry in the IR. This has also the
consequence of providing a correct match between the 4D effective
gauge coupling with the 5D gauge coupling in the limit of
$z_{IR}\rightarrow+\infty$ as was shown in \cite{Cacciapaglia}.  We
therefore write the bulk action for the vector supermultiplet
as~\cite{MartiPomarol},

\begin{eqnarray}
&& S_V =\nonumber\\
&& \int {d^4 xdz \cdot \frac{R}{z}\left\{ {\frac{1}{4}\int {d^2
\theta }  W_\alpha  W^\alpha \Phi +h.c. + \frac{1}{2}\int {d^4
\theta  \left( {\partial _z V - \frac{R}{z}\frac{{(\sigma  +
\sigma ^\dag  )}}{{\sqrt 2 }}} \right)^2 } } \left(\Phi +
\Phi^\dag\right)\right\}}\, ,\nonumber\\
\end{eqnarray}

\noindent and assume that the dilaton superfield acquires a VEV
in its scalar component, $\left\langle
\Phi \right\rangle = e^{ - 2 u z} / g_5^2$. In order to obtain the
same expressions as those found in \cite{PomarolGherghetta}, we
rescale the fields as, $ A_5 \to \frac{z}{R}A_5 $, $\lambda _1 \to
\left( {\frac{R}{z}} \right)^{3/2} \lambda _1 $ and $\lambda _2
\to  i \left( {\frac{R}{z}} \right)^{1/2} \lambda _2 $ and get the
bulk action in components. The action for the bosonic component
fields $(A_M,\Sigma, D)$ is given by,
\begin{eqnarray}
 S_B  &=& \int {d^5 x} \frac{R}{z} \cdot \frac{e^{ - 2 u z}}{ g_5^2 } \cdot \left( - \frac{1}{{4 }}F_{\mu \nu } F^{\mu \nu }
 - \frac{1}{{2 }}\left( {\partial _z A_\mu  } \right)^2
 + \partial _z A_\mu  \partial ^\mu  A_5 - \frac{1}{{2}}\left( {\partial _\mu  A_5 } \right)^2  \right.\nonumber \\
 &+& \left. \frac{1}{2} \left( \frac{R}{z} \right )^2 \cdot \Sigma \partial _\mu  \partial ^\mu  \Sigma
 - \frac{R}{z}\cdot \Sigma \partial _z D + \frac{1}{2}D^2 + \bar{F}_{\sigma} F_{\sigma}\left(\frac{R}{z}\right)^2 \right)
\label{boson} \end{eqnarray} As can be seen in Eq.~(\ref{boson}),
there is a mixing term between $A_{\mu}$ and $A_5$. We can add a
gauge fixing term to remove such mixing,
\begin{eqnarray}
S_{GF}  =  - \int {d^5 x\frac{R}{z} \cdot\frac{e^{ - 2 u z}}{
g_5^2 } \cdot \frac{1}{2}} \left(\partial _\mu  A^\mu   +
\frac{z}{R}\partial _z \left( \frac{R}{z} A_5\right)- 2 u \; A_5
\right)^2
\end{eqnarray}
which corresponds to the unitary gauge in our model. Substituting
$ \Phi$ by its VEV in the lagrangian, we find that the
longitudinal components of the excited gauge bosons states
$A_{\mu}$ are related to the scalar field $A_5$ in the following
way\footnote{This would correspond to the longitudinal modes $A_5$
which are eaten by the $A_{\mu}$ KK gauge bosons.}:
\begin{equation}
p^\mu  A_\mu(p,z)  + \left(\partial _z - \frac{1}{z} -2 u \right)
A_5(p,z) =0.
\end{equation}

The auxiliary fields $F_{\sigma}$ and $D$ can be integrated out
using their equations of motion,
\begin{equation}
F_{\sigma} = 0, \qquad  \qquad D =  - \frac{R}{z}\left(\partial _z
- \frac{2}{z}- 2u \right)\Sigma\,.
\end{equation}

We would like to concentrate on the gaugino sector and study the
effects of SUSY breaking on the spectrum. For that purpose, we
calculate the gaugino bulk action,
\begin{eqnarray}
S_F  &=& \int {d^5 x} \left( {\frac{R}{z}} \right)^4 \cdot
\frac{e^{ - 2 u z}}{ g_5^2 } \cdot \left[ - i\lambda _1 \sigma
^\mu  \partial _\mu \bar \lambda _1 - i\lambda _2 \sigma ^\mu
\partial _\mu  \bar \lambda _2  + \frac{1}{2}\lambda _2 \overleftrightarrow {\partial_z} \lambda _1 \right.\nonumber \\ &-& \left.\frac{1}{2}\bar \lambda _1
\overleftrightarrow {\partial_z} \bar \lambda _2 +
\frac{1}{z}\left( \frac{1}{2} +  u z \right)\lambda _1 \lambda _2
+ \frac{1}{z}\left( \frac{1}{2} +u z  \right) \bar \lambda _1 \bar
\lambda _2 \right] .\label{gauginoaction}
\end{eqnarray}
As can be seen in Eq.~(\ref{gauginoaction}), the dilaton field
modifies the gaugino mass term, introducing a $z$-dependent bulk
mass which leads to a mass gap in the continuum, as we found for
the matter fields. As a matter of fact, this action is exactly
analogous to the fermion matter field action in the case $c=1/2$.

By analogy with the matter fields, we can construct in a simple
manner the solutions to the 5D profiles after rescaling all
component fields by a factor $e^{u z}$,
\begin{eqnarray}
 \lambda_1 (p,z) & = &  \chi _{\rm{4}} (p) e^{uz} \left(\frac{z}{z_{UV}}\right)^2 h_L , \quad
 A_\mu  (p,z) = A_{\mu 4 } (p) e^{uz}  \left( \frac{z}{z_{UV}} \right)^{1/2} h_L ,\\
 \lambda_2 (p,z) & = &  \psi _{\rm{4}} (p)e^{uz} \left(\frac{z}{z_{UV}}\right)^2 h_R, \quad
\Sigma =  \phi _4 (p)e^{uz} \left( \frac{z}{z_{UV}} \right)^{3/2}
h_R\, ,
\end{eqnarray}
where $h_{L,R}$ represents $f_{L,R}$ evaluated at $c=1/2$.  In the
following, we assume that $\sigma(z_{UV})=0$. In this case the
vector superfield acts as the source of a vector superfield CFT
operator with canonical dimensions~\cite{Cacciapaglia:2008bi}.

We study how SUSY breaking affects the super-vector CFT operator
using the holographic language. For this purpose, we add a
Majorana mass term for the $\lambda_{1}$ gauginos on the UV brane,

\begin{equation}
\delta S = \int {d^4 x}\left( \frac{R}{z_{UV}} \right )^4 \int dz
(m z_{UV}   \lambda_1   \lambda_1+h.c. ) \delta (z - z_{UV}).
\end{equation}

In analogy with the matter field case, we can calculate the
kinetic term for the $\lambda_{2}$ gauginos, which are related to
the correlator of the fermionic CFT operator
$\Delta_{f}=-\Sigma_{\lambda_{2}}$, as:
\begin{equation}
\Sigma_{\lambda_2}(p^2)=\left(\frac{R}{z_{UV}}\right)^4
\frac{h_{L}}{h_{R}- m z_{UV} h_{L}} \frac{p_\mu\sigma^\mu}{p}.
\end{equation}
In the conformal limit $\epsilon\rightarrow 0$, the correlator
takes the form,
\begin{equation}
\Delta_{f}\approx - \frac{1}{\gamma \, p^2 + p^2 \ln [2\sqrt {u^2
- p^2 }\, \epsilon ]
 - \sqrt{p^2}\,m}\,\, p_\mu\sigma^\mu\, ,
\end{equation}
where the pole, in the case $p\ll\mu$, is localized  at,
\begin{equation}
p^2_{pole}=\frac{m^2}{\left(\gamma + \ln [ 2 \, u \, \epsilon
]\right)^2}.
\end{equation}
In the case $m^2=0$ the pole is localized at zero momenta as
expected. The continuum spectra starts at momenta $p>u$.  As an
example, for $u=1$ TeV, $\epsilon=10^{-19}$ $\rm{GeV}^{-1}$ and
$m=40$ TeV,  the pole merges with the continuum.

%%%%%%%%%%%%%%%%%%%%%%%%%%%%%%%%%%%%%%%%%%%%%%%%%%%%%%
%%%%%%%%%%%%%%%%%%%%%%%%%%%%%%%%%%%%%%%%%%%%%%%%%%%%%%

%%%%%%%%%%%%%%%%%%%%%%%%%%
\section{Phenomenology and Conclusions}

In this paper we discussed a novel possibility for  supersymmetric
extensions of the Standard Model, where there are continuum excitations
of the SM fields and their superpartners arising
from conformal dynamics. Using the AdS/CFT correspondence we
can explicitly construct the models and explore the properties
of the continuum states. In the supersymmetric limit, the SM
particles and their superpartners are zero modes in the
5D theory, and there is a continuum excitation for each of them
starting at some mass gap due to conformal breaking in the infrared.
After SUSY breaking, the zero-mode superpartners acquire
SUSY-breaking masses and are lifted from the massless spectrum.
For large enough SUSY breaking, the superpartner may merge into
the continuum and there is no longer a well-defined superpartner
state of a definite mass. In the simple setup considered in this paper
where SUSY breaking is localized on UV brane,
the mass gap for the continuum governed by the infrared conformal breaking
does not  change due to locality in the extra dimension, but the shape of the
spectral density is modified by SUSY breaking. One can
imagine that in a more general setup, the continuum excitations
of the SM particles and their superpartners will have different mass gaps.

As we have seen in the discussions of the previous sections, the
properties of superpartners can be quite different in this type of
model, depending on the parameters. The superpartner of a SM
particle could be either a discrete mode below a continuum, the
first of a series of discrete modes, or just a continuum. The last
case will of course be the most challenging to uncover
experimentally.  At the LHC we would expect that most of the time
the superpartner is produced near the bottom of the continuum due
to the fall off in parton distribution functions.  It will be
difficult to construct any peak or edge since there is an
additional smearing of the mass by the continuous spectrum itself.
If the superpartner is produced well above the threshold, then
there is also the possibility of extended decay chains.  This
arises when the superpartner decays to another state, if the new
state is above the original threshold, then it can decay back to
the original superpartner, just at a lower point in the continuum
spectrum.  These events are expected to have large multiplicities
and more spherical shapes, as a reflection of the underlying
conformal theory \cite{Spherical}. The collider phenomenology of
such models is currently under investigation. Serious work will
need to be done to extend current LHC analysis to cover this type
of new physics.

%%%%%%%%%%%%%%%%%%%%%%%%%%
%\section{Conclusion}

%We have seen how SUSY unparticles can be constructed using the
%AdS/CFT correspondence, and that there are three possibilities for
%the spectrum; a continuum without a gap, a single discrete mode
%below a continuum, a hydrogen-like spectrum below a continuum.  In
%the simplest case introducing SUSY breaking can raise the discrete
%modes, and can even move the discrete mode entirely into the
%continuum.  Applying these scenarios as extensions of the MSSM
%gives us startlingly new possibilities for LHC phenomenology.
%Superpartners of ordinary particles might only be a continuum
%rather than a discrete state.  Such continuum superpartners can
%decay to other states and then back to themselves at a lower point
%in the continuum.  Further study is needed to understand how to
%unravel such signals at the LHC.

%%%%%%%%%%%%%%%%%%%%%%%%%%
\section*{Acknowledgments}
We thank Ben Allanach, Jon Bagger, Csaba Csaki,
Adam Falkowski, Markus Luty, Nathan Seiberg, Yuri Shirman and Matt Strassler for
useful discussions and comments. The authors are supported by the
US department of Energy under contract DE-FG02-91ER406746. HC and JT acknowledge
the hospitality of Aspen Center for Physics where part of this work was done.

%%%%%%%%%%%%%%%%%%%%%%%%
\section*{Appendix: Wavefunctions}
We now obtain the spectrum by solving for the wavefunctions of the modes.  As
mentioned before in the text, the general solution to the EOM is a
linear combination of Whittaker functions of the first and second
order. Now if we consider a left-handed  zero mode which
therefore has a Neumann (even) boundary condition at the UV
brane, the accompanying right-handed solution has a
Dirichlet (odd) boundary condition at the same brane. Furthermore,
for momenta smaller than the mass gap, the wavefunction is
normalizable in the sense that it is squared integrable when the
IR branes is taken to infinity. Thus we have,
\begin{equation}
f_R(p,\epsilon)=0, \quad\quad
\int^{\infty}_{\epsilon}\left(\frac{R}{z}\right)^\beta|\left(\frac{z}{\epsilon}\right)^\alpha
f_R(p,z)|^2\frac{1}{\epsilon}\;dz=1\, ,\label{BCN}
\end{equation}
where for fermions $\beta=4$ and $\alpha=2$, and for scalars
$\beta=3$ and $\alpha=3/2$. These two conditions completely fix the
solution to the EOM and the spectrum for momenta smaller than the
mass gap.

There are two independent solutions to the second order equations
of motion, $M(\kappa, m, \zeta)$ and $M(\kappa, -m, \zeta)$. These two
solutions tend to diverge as $z\rightarrow +\infty$. However one
can construct a linear combination of them which is exponentially
decaying as $z\rightarrow +\infty$. This linear combination turns
out to be the Whittaker function of the second kind, $W(\kappa, m, \zeta)$,
defined through Eq.~(\ref{Wdef}). The condition of normalizability
forces us to drop the divergent component $M(\kappa ,m, \zeta)$ in
Eq.~(\ref{fr}) and only keep $W(\kappa, m, \zeta)$.

Let us analyze how the expected massless pole arises by inspecting
Eq.~(\ref{fr}) in the limit $p\rightarrow 0$. From the solution to
Eq.~(\ref{1storder1}), which is always normalizable for $\mu>0$,
independent of the value of $c$, we expect to always find a
solution to Eq.~(\ref{BCN}) for vanishing momentum, $p$, and
positive $\mu$. This in fact, turns out to be always the case
since the pre-factor $p/(\mu+\sqrt{\mu^2-p^2})$ in the second term
of Eq.~(\ref{fr}) vanishes at $p=0$ and furthermore for
non-vanishing $\epsilon$, $W(-c,-1/2+c,2|\mu|\epsilon)$ is regular
and non-zero for any value of $c$. This important pre-factor
arises when relating the normalizable solutions (for $p^2<\mu^2$) to
Eqs.(\ref{2ndorder1}--\ref{2ndorder2}) by
Eqs.(\ref{1storder1}--\ref{1storder2}). Thus, we see that the
appearance of a massless mode, independent of the value of $c$, is
very different from the previously gapless ($\mu=0$) cases
analyzed in the literature
\cite{Cacciapaglia:2008bi,Contino:2004vy} where massless
modes arise only for certain values of $c$. We thus expect that
the corresponding CFT operator will always have a
massless chiral mode for non-zero $\mu$.  Since
Eqs.(\ref{1storder1}--\ref{1storder2}) are symmetric under the
exchange $m(z)\rightarrow -m(z)$ and $f_{R}\rightarrow -f_{L}$ we see that flipping the sign
of  $\mu$ simply flips the handedness of the zero-mode.

Now, let us concentrate for a moment on non-vanishing momenta and
demand that we satisfy the UV boundary condition,
$f_R(p,\epsilon)=0$. From the last discussion we realize we need
to demand,
\begin{equation}
W(-\frac{c\mu}{\sqrt{\mu^2-p^2}},1/2-c,2\sqrt{\mu^2-p^2}\epsilon)=0\,.
\end{equation}
To analyze this last expression, let us expand for small $\zeta$
Eq.~(\ref{Wdef}) using Eq.~(\ref{Mdef}), $M(\kappa,m,\zeta)\approx
\zeta^{1/2+m}$,
\begin{equation}
W(\kappa,m,\zeta)\approx\frac{\Gamma(-2m)}{\Gamma(1/2-m-\kappa)}\zeta^{1/2+m}+\frac{\Gamma(2m)}{\Gamma(1/2+m-\kappa)}\zeta^{1/2-m}.\label{Wapprox}
\end{equation}
In our case, $m=1/2-c$, $\kappa=-c\mu/\sqrt{\mu^2-p^2}$ and
$\zeta=2\sqrt{\mu^2-p^2} z$. The two terms then can compensate
each other and we find that the following relation is satisfied,
\beq \frac{{\Gamma ( - 1 + 2c)}}{{\Gamma (c + \frac{{c\mu}}{{\sqrt
{\mu^2  - p^2 } }})}}2^{1 - 2c} \left(\sqrt {\mu^2 - p^2 } \epsilon \right)^{1
- 2c}  + \frac{{\Gamma (1 - 2c)}}{{\Gamma (1 - c +
\frac{{c\mu}}{{\sqrt {\mu^2  - p^2 } }})}}   = 0. \label{poles}\eeq
We are interested in range $-1/2<c<1/2$, so we see
that the first term in Eq.~(\ref{poles})
vanishes as $\epsilon\rightarrow 0$.
Thus, in order to satisfy the UV boundary condition, we have,
\begin{equation}
1-c+\frac{c\mu}{\sqrt{\mu^2-p^2}}=-n+\delta\;,n\in\mathbb{Z}_0^+\, ,\label{resonances}
\end{equation}
which can only be satisfied for $c<0$.
We can approximately solve for $\delta$, as the solution of,
\begin{equation} \frac{{\Gamma ( - 1 + 2c)}}{{\Gamma ( - 1 + 2c - n)}}2^{1 - c} \left( {\frac{{c\mu}}{{ - 1 + c - n}}z_{UV} } \right)^{1 - c}  +
\frac{{\Gamma (1 - 2c)}}{{\Gamma ( - n + \delta )}}2^c \left(
{\frac{{c\mu}}{{ - 1 + c - n}}z_{UV} } \right)^c  = 0 \end{equation}
and find that,
\begin{equation} \delta\approx (\epsilon \mu)^{1-2c} \frac{(-1)^{n}  (2 c/(c-n-1))^{1-2c} \Gamma (-1+2c)}{ \Gamma (1-2c) \Gamma (-1+2c-n) \Gamma (1+n)},
   \end{equation}
which is tiny in the limit $\epsilon\rightarrow 0$. As an example, in
the case of $c=-0.2$, $\mu=200$ GeV,
$\epsilon=10^{-19}\;\rm{GeV}^{-1}$ and $n=1$, we find
$\delta\approx 10^{-24}$, so we can safely neglect this
correction.

%
%\begin{equation}
%\frac{2^{-c}\left(\rm{Abs}\left[\frac{c\mu}{(1-c+n-\delta)}\right]\epsilon\right)^c\Gamma(1-2c)}{
%\Gamma\left(1-c+\frac{c\mu|1-c+n-\delta|}{|c\mu|}\right)}+\frac{2^{1+c}\left(\rm{Abs}\left[\frac{c\mu
%}{1-c+n-\delta}\right]\epsilon\right)^{1+c}\Gamma(-1+2c)} {\Gamma\left(c+\frac{c\mu|1-c+n-\delta|}{|c\mu|}\right)}=0
%\end{equation}
%

Neglecting $\delta$, we re-write Eq.~(\ref{resonances}) as
\begin{equation}
\sqrt{\mu^2-p^2}=\frac{c\,\mu}{c-1-n}.
\end{equation}
So we require  that $\mu>0$ for this expression to be positive  (with $-1/2<~c~<0$). Solving for $p^2$ we obtain,
\begin{equation}
p^2=\left[ 1-\frac{c^2}{\left(1-c+n\right)^2} \right] \mu^2 .\label{p2}
\end{equation}
Analyzing this solution, we find that most resonances are
localized near $p\sim\mu$ where they accumulate. We can see this
by taking the limit $n\rightarrow +\infty$ in Eq.~(\ref{p2}) and
find that $p^2\rightarrow\mu^2$.

\end{document}